\documentclass[pre, preprint]{revtex4}
\usepackage{amsfonts}
\usepackage{amsmath}
\usepackage{amssymb}
\usepackage{graphicx}
\usepackage{epsf}

\begin{document}

\title{Distance of closest approach of two arbitrary hard ellipses in 2D}
\author{Xiaoyu Zheng}
\affiliation{Department of Mathematical Sciences, Kent State University}
\author{Peter Palffy-Muhoray}
\affiliation{Liquid Crystal Institute, Kent State University}

\begin{abstract}
The distance of closest approach of hard particles is a key parameter of
their interaction and plays an important role in the resulting phase
behavior. For non-spherical particles, the distance of closest approach
depends on orientation, and its calculation is surprisingly difficult.
Although overlap criteria have been developed for use in computer
simulations \citep{Vieillard-baron72,
Perram&W85}, no analytic solutions have been obtained for the distance of
closest approach of ellipsoids in 3-D, or, until now, for ellipses in 2-D.
We have derived an analytic expression for the distance of closest approach
of the centers of two arbitrary hard ellipses as function of their
orientation relative to the line joining their centers. We describe our
method for solving this problem, illustrate our result, and discuss its
usefulness in modeling and simulating systems of anisometric particles such
as liquid crystals.
\end{abstract}

\maketitle

\section{Introduction}

Short range repulsive forces between atoms and molecules in soft condensed
matter are often modeled by an effective hard core, which governs the
proximity of neighbors. Since the attractive interaction with a few nearest
neighbors usually dominates the potential energy, the distance of closest
approach is a key parameter in statistical descriptions of condensed phases.
Simple atoms and molecules with spherical symmetry can be viewed as having
spherical hard cores; the distance of closest approach of the centers of
identical hard spheres in 3-D or of hard circles in 2-D is the diameter. For
non-spherical molecules, such as the constituents of liquid crystals, the
distance depends on orientation, and its calculation is surprisingly
difficult \cite{problem}. The simplest smooth non-spherical shapes are the
ellipse and the ellipsoid. Although overlap criteria have been developed for
use in computer simulations \citep{Vieillard-baron72, Perram&W85}, no
analytic solutions for the distance of closest approach have been obtained
for ellipsoids in 3-D, or, up to now, for ellipses in 2-D. The problem of
determining the distance of closest approach for two ellipses is
particularly intriguing because of its seductive apparent simplicity \cite%
{problem}. We have recently succeeded in deriving an analytic expression for
the distance of closest approach of the centers of two arbitrary hard
ellipses as function of their orientation relative to the line joining their
centers. We describe our method for solving this problem, give the solution,
illustrate our results, and discuss its usefulness in modeling and
simulating systems of anisometric particles such as liquid crystals.

\section{Statement of the problem}

\begin{figure}[ht]
\hbox{\hskip 0cm\epsfxsize14cm \epsfysize10cm
\epsffile{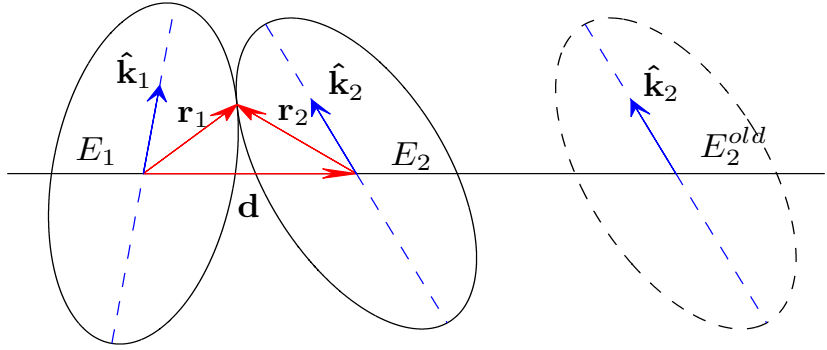}}
\caption{Two externally tangent ellipses $E_1$ and $E_2$. The directions of
the major axes are given by ${\mathbf{\hat{k}}}_1$ and ${\mathbf{\hat{k}}}_2$%
, the centers are joined by the vector {${\mathbf{d}}$}.}
\label{fig-ellipse}
\end{figure}

We consider two ellipses $E_{1}$ and $E_{2}$ in 2D with semi-axes lengths $%
a_{i}$ and $b_{i}\,$where $a_{i}>b_{i}$, eccentricity $e_{i}=\sqrt{1-\dfrac{%
b_i^2}{a_i^2}}$, and major axes oriented along the unit vectors ${\mathbf{%
\hat{k}}}_{i}\,(i=1,2)$. Initially the ellipses are distant so that they
have no point in common. One ellipse is then translated towards the other
along the line joining their centers until they are in point contact
externally (see Fig.\ \ref{fig-ellipse}). The problem is to find the
distance $d$ between centers when the ellipses are so tangent; that is, to
find the distance of closest approach.

The equation of the ellipses are:
\begin{equation}  \label{eq-ellipses}
E_{1}:\quad {\mathbf{r}}_{1}{\mathbb{A}}_{1}{\mathbf{r}}_{1}=1,\quad {%
\mathbb{A}}_{1}=\dfrac{1}{b_{1}^{2}}\left( {\mathbf{I}}+\left( \dfrac{%
b_{1}^{2}}{a_{1}^{2}}-1\right) {\mathbf{\hat{k}}}_{1}{\mathbf{\hat{k}}}%
_{1}\right) =\dfrac{1}{b_{1}^{2}}({\mathbf{I}}-e_{1}^{2}{\mathbf{\hat{k}}}%
_{1}{\mathbf{\hat{k}}}_{1})
\end{equation}%
and%
\begin{equation}
E_{2}:\quad {\mathbf{r}}_{2}{\mathbb{A}}_{2}{\mathbf{r}}_{2}=1,\quad {%
\mathbb{A}}_{2}=\dfrac{1}{b_{2}^{2}}\left( {\mathbf{I}}+\left( \dfrac{%
b_{2}^{2}}{a_{2}^{2}}-1\right) {\mathbf{\hat{k}}}_{2}{\mathbf{\hat{k}}}%
_{2}\right) =\dfrac{1}{b_{2}^{2}}({\mathbf{I}}-e_{2}^{2}{\mathbf{\hat{k}}}%
_{2}{\mathbf{\hat{k}}}_{2}),
\end{equation}%
where ${\mathbf{I}}$ is the identity matrix and ${\mathbf{\hat{k}}}_{i}{%
\mathbf{\hat{k}}}_{i}$ is the dyad product. The vector joining the centers
is given by ${\mathbf{d}}=d\mathbf{\hat{d};}$ $\mathbf{\hat{d}}$ is a given
unit vector. Our goal is to find the distance $d$ as function of ellipse
parameters $a_{1},b_{1},a_{2},b_{2}$ and orientations ${\mathbf{\hat{k}}}%
_{1}\cdot \mathbf{\hat{d}}$, ${\mathbf{\hat{k}}}_{2}\cdot \mathbf{\hat{d}}$
and ${\mathbf{\hat{k}}}_{1}\cdot {\mathbf{\hat{k}}}_{2}$.

It is tempting to seek a solution by solving the quadratic equations

\begin{equation}
{\mathbf{r}}_{1}{\mathbb{A}}_{1}{\mathbf{r}}_{1}=1
\end{equation}%
and%
\begin{equation}
(\mathbf{r}_1-\mathbf{d})\mathbb{A}_2(\mathbf{r}_{1}-\mathbf{d})=1
\end{equation}%
simultaneously for the points of intersection, and then requiring that the
distance $d$ between centers be such that there is intersection exactly at
one point. This approach fails for the following reason: although the
components of $\mathbf{r}_{1}$ at the points of intersection can be obtained
by solving a quartic equation (say for the $x$-component of $\mathbf{r}_{1}$%
), the condition requiring that the quartic have exactly one double real
root is not straightforward to implement (there are four roots, and it is
not clear which two roots need to coalesce to yield the required tangency
condition) and it further gives an equation in $d$ whose order is higher
than quartic, and which cannot therefore be solved analytically.

\section{The Solution}

\label{Sec-problem}

Our approach proceeds via three steps:

(1) Transformation of the two tangent ellipses $E_{1}$ and $E_{2}$, whose
centers are joined by the vector ${\mathbf{d}}$, into a circle $%
C_{1}^{\prime}$ and an ellipse $E_{2}^{\prime}$, whose centers are joined by
the vector ${\mathbf{d}}^{\prime}$. The circle $C_{1}^{\prime}$ and the
ellipse $E_{2}^{\prime}$ remain tangent after the transformation.

(2) Determination of the distance $d^{\prime}$ of closest approach of $%
C_{1}^{\prime}$ and $E_{2}^{\prime}$ analytically.

(3) Determination of the distance $d$ of closest approach of $E_{1}$ and $%
E_{2}$ by inverse transformation of the vector ${\mathbf{d}}^{\prime}$.

\subsection{Transformations}

\begin{figure}[ht]
\hbox{\hskip 1cm\epsfxsize10cm \epsfysize8cm \epsffile{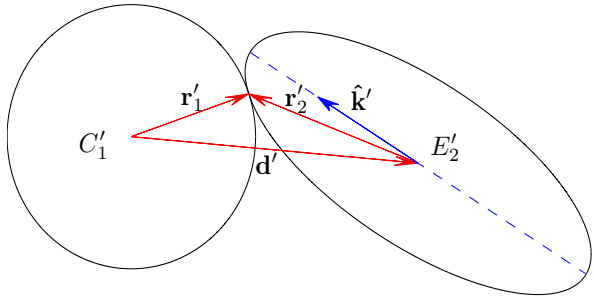}}
\caption{The transformed circle $C^{\prime }_1$ and ellipse $E^{\prime }_2$
after applying transformation $\mathbb{T}$ to the ellipses in Fig.\ \protect
\ref{fig-ellipse}. The major axis of transformed ellipse $E^{\prime }_2$ is
along ${\mathbf{\hat{k}}}^{\prime }$, and the centers are joined by the
vector ${\mathbf{d}}^{\prime }$. }
\label{fig-circle}
\end{figure}

An ellipse can be transformed into a unit circle by anisotropic scaling. We
introduce for this purpose the matrix $\mathbb{T}$, which transforms the
ellipse $E_{1}$ into a unit circle $C_{1}^{\prime }$ and the ellipse $E_{2}$
into another ellipse $E_{2}^{\prime }$. The transformation is a scaling by
the factor $1/a_{1}$ along the $\mathbf{\hat{k}}_{1}$ direction and by the
factor $1/b_{1}$ in the direction perpendicular to $\mathbf{\hat{k}}_{1}$.
The transformation matrix ${\mathbb{T}}$, which transforms position ${%
\mathbf{r}}$ to a position ${\mathbf{r}}^{\prime}$ in a space with
dimensionless coordinates, is
\begin{equation}
{\mathbb{T}}=\dfrac{1}{b_{1}}\left( {\mathbf{I}}+\left( \dfrac{b_{1}}{a_{1}}%
-1\right) {\mathbf{\hat{k}}}_{1}{\mathbf{\hat{k}}}_{1}\right)
\label{transfomationT}
\end{equation}%
and the inverse, ${\mathbb{T}}^{-1}$, is
\begin{equation}
{\mathbb{T}}^{-1}={b_{1}}\left( {\mathbf{I}}+\eta {\mathbf{\hat{k}}}_{1}{%
\mathbf{\hat{k}}}_{1}\right),  \label{transformationT-1}
\end{equation}%
where
\begin{equation}
\eta =\left( \dfrac{a_{1}}{b_{1}}-1\right).
\end{equation}%
One can easily verify that $\mathbb{T}^{-1}{\mathbb{A}}_{1}\mathbb{T}^{-1}={%
\mathbf{I}}$.

If ${\mathbf{r}}_{i}^{\prime }={\mathbb{T}}{\mathbf{r}}_{i}$, or,
equivalently, ${\mathbf{r}}_{i}={\mathbb{T}}^{-1}{\mathbf{r}}_{i}^{\prime }$
$(i=1,2)$, substitution into Eq.\ (\ref{eq-ellipses}) gives a unit circle $%
C_{1}^{\prime }$ and a new ellipse $E_{2}^{\prime }$ (see Fig.\ \ref%
{fig-circle}). That is,
\begin{equation}
C_{1}^{\prime }:\quad {\mathbf{r}}_{1}{\mathbb{A}}_{1}{\mathbf{r}}_{1}={%
\mathbf{r}}_{1}^{\prime }{\mathbb{T}}^{-1}{\mathbb{A}}_{1}{\mathbb{T}}^{-1}{%
\mathbf{r}}_{1}^{\prime }={\mathbf{r}}_{1}^{\prime }{\mathbf{r}}_{1}^{\prime
}=1
\end{equation}%
and%
\begin{equation}
E_{2}^{\prime }:\quad {\mathbf{r}}_{2}{\mathbb{A}}_{2}{\mathbf{r}}_{2}={%
\mathbf{r}}_{2}^{\prime }{\mathbb{T}}^{-1}{\mathbb{A}}_{2}{\mathbb{T}}^{-1}{%
\mathbf{r}}_{2}^{\prime }={\mathbf{r}}_{2}^{\prime }{\mathbb{A}}^{\prime }{%
\mathbf{r}}_{2}^{\prime }=1.
\end{equation}%
${\mathbb{A}}^{\prime }$ can be written as

\begin{equation}
{\mathbb{A}}^{\prime }=\dfrac{b_{1}^{2}}{b_{2}^{2}}({\mathbf{I}}+\eta {%
\mathbf{\hat{k}}}_{1}{\mathbf{\hat{k}}}_{1})({\mathbf{I}}-e_{2}^{2}{\mathbf{%
\hat{k}}}_{2}{\mathbf{\hat{k}}}_{2})({\mathbf{I}}+\eta {\mathbf{\hat{k}}}_{1}%
{\mathbf{\hat{k}}}_{1}).
\end{equation}

The eigenvectors of ${\mathbb{A}}^{\prime }$ provide information about the
directions of the principal axes and the eigenvalues about the lengths of
the semi-axes of the transformed ellipse $E_{2}^{\prime }$. Since ${\mathbb{A%
}}^{\prime }$ is real symmetric, its eigenvalues $\lambda _{+}$ and $\lambda
_{-}$ are real, and the corresponding eigenvectors $\mathbf{\hat{k}}%
_{+}^{\prime }$ and $\mathbf{\hat{k}}_{-}^{\prime}$ are orthogonal. Explicit
expressions for these are given in the Appendix. The lengths of the
semi-axes, if $\lambda _{+}>\lambda _{-}$, are given by
\begin{equation}
\begin{split}
b_{2}^{\prime }& =\frac{1}{\sqrt{\lambda _{+}}}, \\
a_{2}^{\prime }& =\dfrac{1}{\sqrt{\lambda _{-}}},
\end{split}%
\end{equation}%
and we note that $a_{2}^{\prime }>b_{2}^{\prime }$.

Under the transformation ${\mathbb{T}}$, the vector ${\mathbf{d}}$ is
transformed to%
\begin{equation}
{\mathbf{d}}^{\prime }={\mathbb{T}}{\mathbf{d}}=d{\mathbb{T}}\mathbf{\hat{d}}%
=d^{\prime }\mathbf{\hat{d}}^{\prime },
\end{equation}%
where $\mathbf{\hat{d}}^{\prime }=\dfrac{{\mathbb{T}}\mathbf{\hat{d}}}{|{%
\mathbb{T}}\mathbf{\hat{d}}|}$ is a unit vector. Explicitly,
\begin{equation}
{\mathbb{T}}\mathbf{\hat{d}=}\dfrac{1}{b_{1}}\left( \mathbf{\hat{d}}+\left(
\dfrac{b_{1}}{a_{1}}-1\right) ({\mathbf{\hat{k}}}_{1}\cdot \mathbf{\hat{d})}{%
\mathbf{\hat{k}}}_{1}\right)
\end{equation}%
and%
\begin{equation}
|{\mathbb{T}}\mathbf{\hat{d}}|=\frac{1}{b_{1}}\sqrt{1-e_{1}^{2}({\mathbf{%
\hat{k}}}_{1}\cdot \mathbf{\hat{d})}^{2}}.
\end{equation}

\subsection{Distance ${\mathbf{d}}^{\prime}$ of closest approach of a circle
and an ellipse}

We next derive the a useful relation between the position vector ${\mathbf{r}%
}$ of a point on the ellipse and the unit outward normal ${{\mathbf{\hat{n}}}%
}$ at that point. For an ellipse, given by ${\mathbf{r}}{\mathbb{A}}{\mathbf{%
r}}=1$, the unit normal ${\mathbf{\hat{n}}} $ is%
\begin{equation}
{\mathbf{\hat{n}}}=\frac{\mathbf{\nabla (}{\mathbf{r}}{\mathbb{A}}{\mathbf{r)%
}}}{|\mathbf{\nabla (}{\mathbf{r}}{\mathbb{A}}{\mathbf{r)}}|}=\dfrac{{%
\mathbb{A}}{\mathbf{r}}}{\sqrt{{\mathbf{r}}{\mathbb{A}}^{2}{\mathbf{r}}}}.
\label{n}
\end{equation}%
Multiplying Eq.\ (\ref{n}) by ${\mathbb{B}}={\mathbb{A}}^{-1}$ gives
\begin{equation}
{\mathbb{B}}{\mathbf{\hat{n}}}=\dfrac{{\mathbf{r}}}{\sqrt{{\mathbf{r}}{%
\mathbb{A}}^{2}{\mathbf{r}}}}  \label{Bn}
\end{equation}%
and multiplying ${\mathbb{B}}{\mathbf{\hat{n}}}$ by ${\mathbf{\hat{n}}}$
gives
\begin{equation}
{\mathbf{\hat{n}}}{\mathbb{B}}{\mathbf{\hat{n}}}=\dfrac{1}{{\mathbf{r}}{%
\mathbb{A}}^{2}{\mathbf{r}}}.  \label{nBn}
\end{equation}%
Substituting into (\ref{Bn}), we obtain ${{\mathbf{r}}}$ in terms of the
unit normal ${\mathbf{\hat{n}}}$
\begin{equation}
{\mathbf{r}}=\dfrac{{\mathbb{B}}{\mathbf{\hat{n}}}}{\sqrt{{\mathbf{\hat{n}}}{%
\mathbb{B}}{\mathbf{\hat{n}}}}}.  \label{rBn}
\end{equation}

If a unit circle and an ellipse are externally tangent, then the directions
of their normals at the point of contact must be opposite. If the unit
outward normal of the unit circle $C_{1}^{\prime }$ at the point of contact
is ${\mathbf{\hat{n}}}^{\prime }$, then
\begin{equation}
\mathbf{r}_{1}^{\prime }={\mathbf{\hat{n}}}^{\prime },\quad \mathbf{r}%
_{2}^{\prime }=-\dfrac{{\mathbb{B}^{\prime }}{\mathbf{\hat{n}}}^{\prime }}{%
\sqrt{{\mathbf{\hat{n}}}^{\prime }{\mathbb{B}^{\prime }}{\mathbf{\hat{n}}}%
^{\prime }}}
\end{equation}%
and we have, for the vector ${\mathbf{d}}^{\prime }$ joining the centers,
\begin{equation}
{\mathbf{d}}^{\prime }=\mathbf{r}_{1}^{\prime }-{\mathbf{r}}_{2}^{\prime }={%
\mathbf{\hat{n}}}^{\prime }+\dfrac{{\mathbb{B}}^{\prime }{\mathbf{\hat{n}}}%
^{\prime }}{\sqrt{{\mathbf{\hat{n}}}^{\prime }{\mathbb{B}}^{\prime }{\mathbf{%
\hat{n}}}^{\prime }}},  \label{rp}
\end{equation}%
where
\begin{equation}
{\mathbb{B}}^{\prime }={{\mathbb{A}}^{\prime }}^{-1}={{b^{\prime }}_{2}^{2}}%
\left( {\mathbf{I}}+\delta {\mathbf{\hat{k}}}_{-}^{\prime }{\mathbf{\hat{k}}}%
_{-}^{\prime }\right)
\end{equation}%
and
\begin{equation}
\delta =\dfrac{{a^{\prime }}_{2}^{2}}{{b^{\prime }}_{2}^{2}}-1>0.
\end{equation}%
Eq. (\ref{rp}) is a key result. It is a vector equation with only two
unknowns: the magnitude of $\mathbf{d}^{\prime }$ and the direction of ${%
\mathbf{\hat{n}}}^{\prime }$. It can be solved for $d^{\prime }$ as follows.
We multiply both sides of Eq.\ (\ref{rp}) by ${\mathbf{\hat{k}}}_{-}^{\prime
}$ and by ${{\mathbf{\hat{k}}}}_{+}^{\prime }$, and letting
\begin{equation}
{\mathbf{\hat{k}}}_{-}^{\prime }\cdot \mathbf{\hat{d}}^{\prime }=\sin \phi
,\quad {\mathbf{\hat{k}}}_{-}^{\prime }\cdot {\mathbf{\hat{n}}}^{\prime
}=\sin \psi ,\quad {{\mathbf{\hat{k}}}}_{+}^{\prime }\cdot \mathbf{\hat{d}}%
^{\prime }=\cos \phi ,\quad {{\mathbf{\hat{k}}}}_{+}^{\prime }\cdot {\mathbf{%
\hat{n}}}^{\prime }=\cos \psi ,
\end{equation}%
we get, from Eq. (\ref{rp}),
\begin{equation}  \label{eq-rphi}
d^{\prime }\sin \phi =\sin \psi \left( 1+\dfrac{b_{2}^{\prime }(1+\delta )}{%
\sqrt{1+\delta \sin ^{2}\psi }}\right)
\end{equation}%
and%
\begin{equation}
d^{\prime }\cos \phi =\cos \psi \left( 1+\dfrac{b_{2}^{\prime }}{\sqrt{%
1+\delta \sin ^{2}\psi }}\right) .
\end{equation}%
Here the unknowns are $\psi $ and $d^{\prime }$. In the special case of $%
\delta =0$, $d^{\prime }=1+b_{2}^{\prime }=1+a_{2}^{\prime }$, and in the
case of $\phi =\dfrac{\pi }{2}$, $d^{\prime }=1+b_{2}^{\prime }\sqrt{%
1+\delta }=1+a_{2}^{\prime }$. In general, $\phi \neq \dfrac{\pi }{2}$, and
the solution for $d^{\prime }$ is more challenging.

We let $q=\sqrt{1+\delta \sin ^{2}\psi }$, then
\begin{equation}
\sin ^{2}\psi =\dfrac{q^{2}-1}{\delta }
\end{equation}
and
\begin{equation}
\cos ^{2}\psi =1-\dfrac{q^{2}-1}{\delta}.
\end{equation}%
Substitution into Eq.\ (\ref{eq-rphi})(a-b), squaring both sides and
dividing these two equations to eliminate $d^{\prime }$ gives a quartic
equation for $q$,
\begin{equation}
\tan ^{2}\phi (\delta +1-q^{2})(\dfrac{q}{b_{2}^{\prime }}+1)^{2}=(q^{2}-1)(%
\dfrac{q}{b_{2}^{\prime }}+1+\delta )^{2}.  \label{quart}
\end{equation}%
This can be written in the standard form $Aq^{4}+Bq^{3}+Cq^{2}+Dq+E=0$,
where the coefficients are
\begin{subequations}
\begin{equation}
A=-\dfrac{1}{{b^{\prime }}_{2}^{2}}(1+\tan ^{2}\phi),
\end{equation}%
\begin{equation}
B=-\dfrac{2}{{b_{2}^{\prime }}}(1+\tan ^{2}\phi +\delta),
\end{equation}%
\begin{equation}
C=-\tan ^{2}\phi -(1+\delta)^{2}+\dfrac{1}{{b^{\prime }}_{2}^{2}}%
(1+(1+\delta )\tan ^{2}\phi),
\end{equation}%
\begin{equation}
D=\dfrac{2}{{b_{2}^{\prime }}}(1+\tan ^{2}\phi )(1+\delta),
\end{equation}%
\begin{equation}
E=(1+\tan ^{2}\phi +\delta )(1+\delta)
\end{equation}%
and
\end{subequations}
\begin{equation}
\tan ^{2}\phi =\frac{({\mathbf{\hat{k}}}_{-}^{\prime }\cdot \mathbf{\hat{d}}%
^{\prime })^{2}}{1-({\mathbf{\hat{k}}}_{-}^{\prime }\cdot \mathbf{\hat{d}}%
^{\prime })^{2}}.
\end{equation}

The roots of Eq. (\ref{quart}) can be obtained explicitly as follows.

To make contact with the standard solution of the quartic equation, using
Ferrari's method \cite{uspensky}, we define
\begin{subequations}
\begin{equation}
\alpha =-\dfrac{3B^{2}}{8A^{2}}+\dfrac{C}{A},
\end{equation}%
\begin{equation}
\beta =\dfrac{B^{3}}{8A^{3}}-\dfrac{BC}{2A^{2}}+\dfrac{D}{A},
\end{equation}%
\begin{equation}
\gamma =\dfrac{-3B^{4}}{256A^{4}}+\dfrac{CB^{2}}{16A^{3}}-\dfrac{BD}{4A^{2}}+%
\dfrac{E}{A}
\end{equation}%
and%
\begin{equation}
P=-\dfrac{\alpha ^{2}}{12}-\gamma,
\end{equation}%
\begin{equation}
Q=-\dfrac{\alpha ^{3}}{108}+\dfrac{\alpha \gamma }{3}-\dfrac{\beta ^{2}}{8}
\end{equation}
and
\begin{equation}
U=\left( -\dfrac{Q}{2}+\sqrt{\dfrac{Q^{2}}{4}+\dfrac{P^{3}}{27}}\right)
^{1/3},
\end{equation}%
where we take the principal values of the roots. If $U=0$, then
\end{subequations}
\begin{equation}
y=-\dfrac{5}{6}\alpha -Q^{1/3},
\end{equation}%
otherwise
\begin{equation}
y=-\dfrac{5}{6}\alpha +U-\dfrac{P}{3U}.
\end{equation}%
In terms of these, the one real positive root $q$ is
\begin{equation}
q=-\dfrac{B}{4A}+\dfrac{1}{2}\left( \sqrt{\alpha +2y}+\sqrt{-\left( 3\alpha
+2y+\dfrac{2\beta }{\sqrt{\alpha +2y}}\right)}\right).
\end{equation}

In the special case when $\alpha +2y=0$, then $\beta =0$ (which we have not
observed in this problem, but include here for completeness) and the
positive real root is given by%
\begin{equation}
q=-\frac{B}{4A}+\sqrt{\frac{-\alpha +\sqrt{\alpha ^{2}-4\gamma }}{2}}.
\end{equation}

Knowing $q$, $d^{\prime }$ can be found by squaring both sides of Eqs.\ (\ref%
{eq-rphi})(a-b) and adding; this gives
\begin{equation}
{d^{\prime }}=\sqrt{\dfrac{q^{2}-1}{\delta }\left( 1+\dfrac{b_{2}^{\prime
}(1+\delta )}{q}\right) ^{2}+\left( 1-\dfrac{q^{2}-1}{\delta }\right) \left(
1+\dfrac{b_{2}^{\prime }}{q}\right) ^{2}}.
\end{equation}%
The vector joining the centers of the circle and the ellipse is given by
\begin{equation}
{{\mathbf{d}}^{\prime }}=d^{\prime }\mathbf{\hat{d}}^{\prime}.
\end{equation}

\subsection{Distance $d$ of closest approach}

The distance of closest approach of the two ellipses is obtained via the
transformation from ${{\mathbf{d}}^{\prime }}$ to ${\mathbf{d}}$,
\begin{equation}
{\mathbf{d}}={\mathbb{T}}^{-1}{{\mathbf{d}}^{\prime }=d}^{\prime }{\mathbb{T}%
}^{-1}\mathbf{\hat{d}}^{\prime }={d}^{\prime }{\mathbb{T}}^{-1}\dfrac{{%
\mathbb{T}}\mathbf{\hat{d}}}{|{\mathbb{T}}\mathbf{\hat{d}}|}=\frac{{d}%
^{\prime }}{|{\mathbb{T}}\mathbf{\hat{d}}|}\mathbf{\hat{d}=}\dfrac{d^{\prime
}}{\dfrac{1}{b_{1}}\sqrt{1-e_{1}^{2}({\mathbf{\hat{k}}}_{1}\cdot \mathbf{%
\hat{d})}^{2}}}\mathbf{\hat{d}},
\end{equation}%
and finally we have%
\begin{equation}
d=\frac{d^{\prime }}{\sqrt{1-e_{1}^{2}({\mathbf{\hat{k}}}_{1}\cdot \mathbf{%
\hat{d})}^{2}}}b_{1}.
\end{equation}%
This is the solution for the distance of closest approach, which is our main
result.

\subsection{Contact point}

In addition to the distance of closest approach, it is interesting and
useful to locate the point contact. We denote the vector from the center of
Ellipse 1 to the point of contact as $\boldsymbol{r}_{c}$. \ In the
transformed coordinate system, where Ellipse 1 has become a unit circle
after the affine transformation, the vector from the center of circle to the
point of contact is $\boldsymbol{r}_{c}^{\prime }={\mathbf{\hat{n}}}^{\prime
}$ where ${\mathbf{\hat{n}}}^{\prime }$ is the unit normal at the point of
contact, whose components along the orthogonal unit vectors ${\mathbf{\hat{k}%
}}_{+}^{\prime }$ and ${{\mathbf{\hat{k}}}}_{-}^{\prime }$ are known \ Thus,
to obtain $\boldsymbol{r}_{c}$, it is only necessary to perform the
transformation of ${\mathbf{\hat{n}}}^{\prime }$, that is,
\begin{equation}
\boldsymbol{r}_{c}={\mathbb{T}}^{-1}{\mathbf{\hat{n}}}^{\prime }
\end{equation}

The components of ${\mathbf{\hat{n}}}^{\prime }$ are given by%
\begin{equation}
{\mathbf{\hat{k}}}_{-}^{\prime }\cdot {\mathbf{\hat{n}}}^{\prime }=\sin \psi
,\quad {{\mathbf{\hat{k}}}}_{+}^{\prime }\cdot {\mathbf{\hat{n}}}^{\prime
}=\cos \psi
\end{equation}%
where $\psi $ is given by
\begin{eqnarray}
\sin \psi &=&sgn(\sin \phi )\sqrt{\frac{q^{2}-1}{\delta }}, \\
\cos \psi &=&sgn(\cos \phi )\sqrt{1-\frac{q^{2}-1}{\delta }}.
\end{eqnarray}%
where $sgn(x)$ gives the sign of $x$ and the angle $\phi $ is known.

It follows that
\begin{equation}
{\mathbf{\hat{n}}}^{\prime }=\cos \psi {{\mathbf{\hat{k}}}}_{+}^{\prime
}+\sin \psi {\mathbf{\hat{k}}}_{-}^{\prime }
\end{equation}%
Writing ${{\mathbf{\hat{k}}}}_{+}^{\prime }$ and ${\mathbf{\hat{k}}}%
_{-}^{\prime }\,$in terms of ${\mathbf{\hat{k}}}_{1}$ and ${\mathbf{\hat{k}}}%
_{2}$
\begin{equation}
{{\mathbf{\hat{k}}}}_{+}^{\prime }=\cos \gamma \frac{(\mathbf{\hat{k}}_{1}+%
\mathbf{\hat{k}}_{2})}{\sqrt{2}\sqrt{1+\mathbf{\hat{k}}_{1}\cdot \mathbf{%
\hat{k}}_{2}}}+\sin \gamma \frac{(\mathbf{\hat{k}}_{1}-\mathbf{\hat{k}}_{2})%
}{\sqrt{2}\sqrt{1-\mathbf{\hat{k}}_{1}\cdot \mathbf{\hat{k}}_{2}}}
\end{equation}%
where%
\begin{equation}
\cos \gamma ={{\mathbf{\hat{k}}}}_{+}^{\prime }\cdot \frac{(\mathbf{\hat{k}}%
_{1}+\mathbf{\hat{k}}_{2})}{\sqrt{2}\sqrt{1+\mathbf{\hat{k}}_{1}\cdot
\mathbf{\hat{k}}_{2}}}
\end{equation}%
and%
\begin{equation}
\sin \gamma ={{\mathbf{\hat{k}}}}_{+}^{\prime }\cdot \frac{(\mathbf{\hat{k}}%
_{1}-\mathbf{\hat{k}}_{2})}{\sqrt{2}\sqrt{1-\mathbf{\hat{k}}_{1}\cdot
\mathbf{\hat{k}}_{2}}}
\end{equation}%
\begin{equation}
{{\mathbf{\hat{k}}}}_{-}^{\prime }=-\sin \gamma \frac{(\mathbf{\hat{k}}_{1}+%
\mathbf{\hat{k}}_{2})}{\sqrt{2}\sqrt{1+\mathbf{\hat{k}}_{1}\cdot \mathbf{%
\hat{k}}_{2}}}+\cos \gamma \frac{(\mathbf{\hat{k}}_{1}-\mathbf{\hat{k}}_{2})%
}{\sqrt{2}\sqrt{1-\mathbf{\hat{k}}_{1}\cdot \mathbf{\hat{k}}_{2}}}
\end{equation}

Substitution gives ${\mathbf{\hat{n}}}^{\prime }$ in terms of ${\mathbf{\hat{%
k}}}_{1}$ and ${\mathbf{\hat{k}}}_{2}$\ gives

\begin{equation}
{\mathbf{\hat{n}}}^{\prime }=\cos (\psi +\gamma )\frac{(\mathbf{\hat{k}}_{1}+%
\mathbf{\hat{k}}_{2})}{\sqrt{2}\sqrt{1+\mathbf{\hat{k}}_{1}\cdot \mathbf{%
\hat{k}}_{2}}}+\sin (\psi +\gamma )\frac{(\mathbf{\hat{k}}_{1}-\mathbf{\hat{k%
}}_{2})}{\sqrt{2}\sqrt{1-\mathbf{\hat{k}}_{1}\cdot \mathbf{\hat{k}}_{2}}}
\end{equation}

Now
\begin{equation}
\boldsymbol{r}_{c}={\mathbb{T}}^{-1}{\mathbf{\hat{n}}}^{\prime }={b{_{1}}%
\left( {\mathbf{I}}+\eta {\mathbf{\hat{k}}}_{1}{\mathbf{\hat{k}}}_{1}\right)
\mathbf{\hat{n}}}^{\prime }
\end{equation}%
and so
\begin{eqnarray}
\boldsymbol{r}_{c} &=&b_{1}\cos (\psi +\gamma )\frac{(\mathbf{\hat{k}}_{1}+%
\mathbf{\hat{k}}_{2})}{\sqrt{2}\sqrt{1+\mathbf{\hat{k}}_{1}\cdot \mathbf{%
\hat{k}}_{2}}}+b_{1}\sin (\psi +\gamma )\frac{(\mathbf{\hat{k}}_{1}-\mathbf{%
\hat{k}}_{2})}{\sqrt{2}\sqrt{1-\mathbf{\hat{k}}_{1}\cdot \mathbf{\hat{k}}_{2}%
}}+  \notag \\
&&(a_{1}-b_{1})\cos (\psi +\gamma )\frac{(1+\mathbf{\hat{k}}_{1}\cdot
\mathbf{\hat{k}}_{2})\mathbf{\hat{k}}_{1}}{\sqrt{2}\sqrt{1+\mathbf{\hat{k}}%
_{1}\cdot \mathbf{\hat{k}}_{2}}}+(a_{1}-b_{1})\sin (\psi +\gamma )\frac{(1-%
\mathbf{\hat{k}}_{1}\cdot \mathbf{\hat{k}}_{2})\mathbf{\hat{k}}_{1}}{\sqrt{2}%
\sqrt{1-\mathbf{\hat{k}}_{1}\cdot \mathbf{\hat{k}}_{2}}}
\end{eqnarray}%
and finally
\begin{eqnarray}
\boldsymbol{r}_{c} &=&[(a_{1}+(a_{1}-b_{1})\mathbf{\hat{k}}_{1}\cdot \mathbf{%
\hat{k}}_{2})\frac{\cos (\psi +\gamma )}{\sqrt{2}\sqrt{1+\mathbf{\hat{k}}%
_{1}\cdot \mathbf{\hat{k}}_{2}}}+(a_{1}-(a_{1}-b_{1})\mathbf{\hat{k}}%
_{1}\cdot \mathbf{\hat{k}}_{2})\frac{\sin (\psi +\gamma )}{\sqrt{2}\sqrt{1-%
\mathbf{\hat{k}}_{1}\cdot \mathbf{\hat{k}}_{2}}}]\mathbf{\hat{k}}_{1}+
\notag \\
&&[b_{1}\cos (\psi +\gamma )\frac{1}{\sqrt{2}\sqrt{1+\mathbf{\hat{k}}%
_{1}\cdot \mathbf{\hat{k}}_{2}}}-b_{1}\sin (\psi +\gamma )\frac{\mathbf{1}}{%
\sqrt{2}\sqrt{1-\mathbf{\hat{k}}_{1}\cdot \mathbf{\hat{k}}_{2}}}]\mathbf{%
\hat{k}}_{2}
\end{eqnarray}

\section{Discussion}

The above methods give a closed form expression for the distance of closest
approach and the position of the point of contact for two ellipses of
arbitrary size, eccentricity and orientation. Detailed steps of the
calculation are given in the Appendix. To demonstrate the applicability of
the method, we give two examples: calculation of the excluded area and the
locus of the point of contact while one ellipse is fixed and the other is
rotated.

\subsection{Excluded area $A_{ex}$}

\begin{figure}[th]
\hbox{\hskip -1cm\epsfxsize10cm \epsfysize8cm
\epsffile{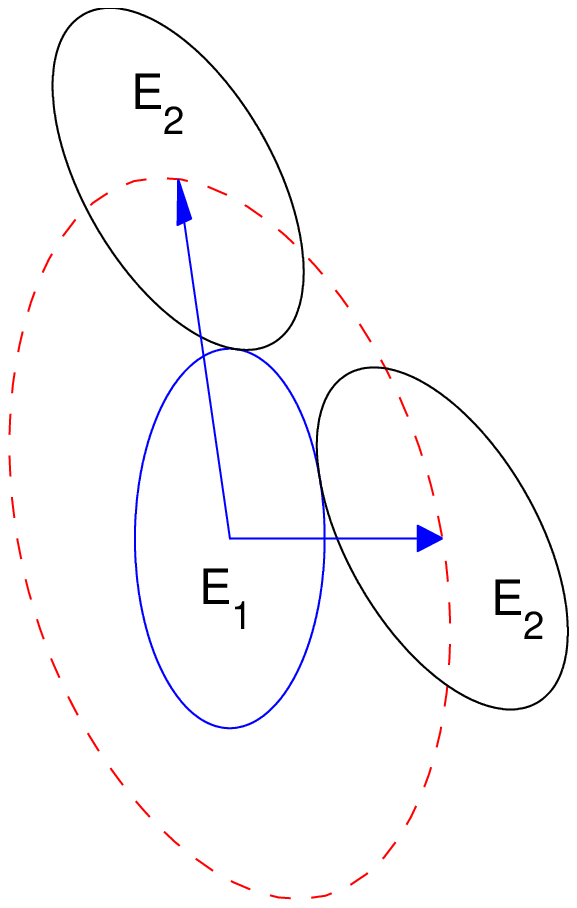}\hskip -2cm\epsfxsize10cm \epsfysize8cm
\epsffile{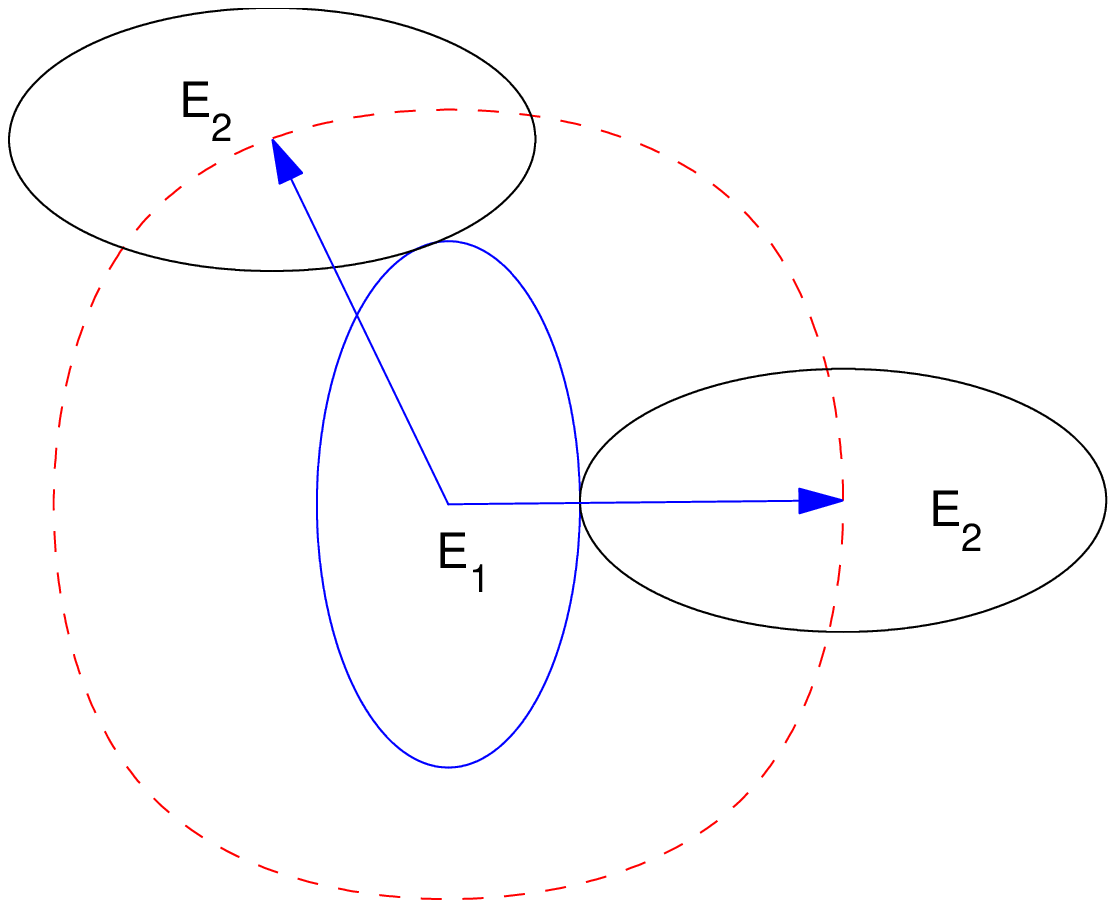}}
\caption{Excluded area for two ellipses. Ellipse $E_{1}$ is fixed at origin,
and ellipse $E_{2}$ rotates around it, keeping its orientation fixed and
remaining tangent to $E_{1}$. The center of $E_{2}$ traces out the dashed
curve. The area bounded by the dashed curve is the excluded area $A_{ex}$.}
\label{fig-excluded}
\end{figure}
From the analytical solution provided in Section \ref{Sec-problem},
one can easily compute, numerically, the excluded area for two
identical ellipses
whose orientation is fixed by integrating $d^{2}(\mathbf{\hat{d}}%
,a_{1},b_{1},a_{2},b_{2},{\mathbf{\hat{k}}}_{1},{\mathbf{\hat{k}}}_{2})$
over $\mathbf{\hat{d};}$%
\begin{equation}
A_{ex}=\frac{1}{2}\int d^{2}(\mathbf{\hat{d}},a_{1},b_{1},a_{2},b_{2},{%
\mathbf{\hat{k}}}_{1},{\mathbf{\hat{k}}}_{2})|d\mathbf{\hat{d}}|.
\end{equation}

Fig.\ \ref{fig-excluded} shows the locus of the center of ellipse $E_{2}$
rotating around $E_{1}$ while keeping the orientation of both ellipses
fixed. Here $a_{1}=a_{2}=2,\,b_{1}=b_{2}=1$. When the angle between the
major axes is $30^{\circ }$, the excluded area is $26.4$ (Fig.\ \ref%
{fig-excluded}.a). If the angle is increased to $45^{\circ }$, then the
excluded area is $27.6$. If the angle is $90^{\circ }$, then the excluded
area is $29.7$ (Fig.\ \ref{fig-excluded}.b). The excluded area increases
monotonically with the angle between major axes of two ellipses; it is the
smallest when the major axes are parallel, and the largest when the major
axes are normal to each other.

\subsection{Locus of the point of contact}

\begin{figure}[ht]
\hbox{\hskip 2cm\epsfxsize10cm \epsfysize8cm \epsffile{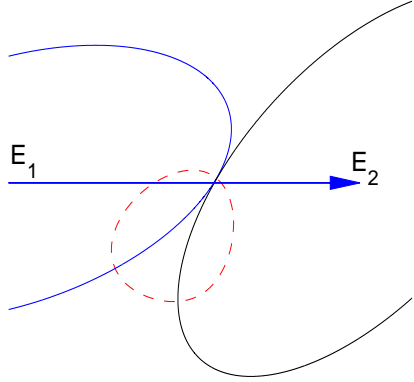}}
\caption{Locus of the point of contact. Ellipse $E_1$ is rotated about its
center, while ellipse $E_2$ keeps its orientation fixed. The center of $E_2$
moves so that $E_2$ remains tangent to $E_1$. The point of contact traces
out the dashed curve. }
\label{fig-locus}
\end{figure}
Fig.\ \ref{fig-locus} shows that locus of the point of contact when ellipse $%
E_{1}$ is rotating about its center while ellipse $E_{2}$ keeps its
orientation. It is interesting and unexpected that the locus has dipolar
rather than quadrupolar symmetry.

\subsection{Potential Applications}

Our result, the analytical expression for the distance of closest approach
of two hard ellipses, has a number of potential applications. It may be
useful in modeling 2-D liquid crystals, both analytically and numerically.
The excluded area, discussed above, is a key parameter in statistical models
\cite{Gelbart77} which can be calculated for ellipses from our result.
Another potential application is in the theory of nematic liquid crystals.
One important contribution to the elastic constants of nematics is due to
anisotropic dispersion forces. The average Van der Waals interaction energy
of a molecule with its neighbors is an algebraic function of the distance of
closest approach. The origin of three distinct elastic constants in nematics
is still unresolved. Our result may be useful in modeling elastic constants
in 2-D nematics, and possibly giving insights towards understanding their
origins in general.

Monte Carlo calculations have played an important role in modeling the phase
behavior of isotropic fluids and liquid crystals \cite{Care&C05}.
Vieillard-Baron developed the first overlap criterion for identical hard
ellipses \cite{Vieillard-baron72}. He derived a contact function $\Psi (a,b,%
\mathbf{\hat{k}}_{1},\mathbf{\hat{k}}_{2},\mathbf{d)}$ such that $\Psi=0 $
when the ellipses are tangent (either exteriorly or interiorly), and this
function is positive and at least one of two auxiliary functions are
negative if the ellipses have no real point in common. This overlap
criterion has been used in Monte-Carlo simulations of hard ellipse systems
\citep{Vieillard-baron72,
Cuesta&F90}. It may be possible to solve $\Psi(a,b,\mathbf{\hat {k}}_{1},%
\mathbf{\hat{k}}_{2},\mathbf{d)}=0$ for $d$ (this involves solving a quartic
equation), and thus obtain a result similar to ours; to our knowledge this
has not yet been done. However, Vieillard-Baron's contact function $\Psi$ is
only valid for identical ellipses, and so this result would not be as
general as ours, presented here.

According to the Hohenberg-Mermin-Wagner theorem, long range order
corresponding to broken continuous symmetry is not allowed in 2D systems
with short-range interactions \cite{HMW}. The possibility of long-range
order in 2D nematics has been discussed theoretically \cite{straley71} and
examined using Monte-Carlo simulations with Lennard-Jones like potentials
\cite{Tobochnik&C83} as well as with hard rods \cite{Bates&F00}. Although it
has been shown that true long range order cannot exist if the interparticle
potential is separable into a positional and an orientational part \cite%
{Tobochnik&C83}, it is not clear what the implications are for systems of
hard ellipses. Frenkel has shown that only quasi-long range order exists for
hard spherocylinders \cite{Bates&F00}, that is, the correlations in
orientational order decay algebraically. Hard ellipsoids, however, can show
dramatically different behavior from hard spherocylinders \cite%
{Allen&E&F&M93} (hard ellipses do not form smectic phases, whereas
spherocylinders do \cite{Wu&S01}), and for this reason Monte Carlo
simulations of hard ellipses, on systems larger than studied by
Vieillard-Baron, would be of considerable interest. Our result for the
distance provides an overlap criterion which could be usefully applied here.

Another area of interest is phase separation in hard particle systems %
\citep{Leptoukh&S&R95, Varga&P&G&F&J05}. For example, simulations of hard
disks and hard parallel squares have been studied, and phase separation has
been observed. Theoretical studies, on the other hand predict no phase
separation in 2D \cite{Perera&C&P&D02}. Our results could provide the
criterion for the overlap of ellipses of different sizes, and thus enable
Monte-Carlo simulations of binary mixtures of hard ellipses.

Vieillard-Baron also provides an overlap criterion for two identical
ellipsoids of revolution in 3D \cite{Vieillard-baron72}. This involves the
evaluation of a contact function $\Psi $ and five auxiliary functions, three
of which must be non-negative and at least one among the remaining three
must be negative to avoid overlap. Perram and Wertheim provided a more
general overlap criterion for hard ellipsoids \cite{Perram&W85}. Their
scheme for evaluating the criterion involves an iterative numerical
technique to find the maximum of a scalar function. Our results can provide
the basis of a simple algorithm to determine the distance of closest
approach of two ellipsoids in 3D. This involves passing a plane through the
line joining the centers of the two ellipsoids, determining the distance of
closest approach of the ellipses in the plane, then rotating the plane and
finding the largest such distance. The details of this algorithm will be
published elsewhere.

\section{Conclusion}

\label{Sec-Conclusions} We have derived an analytic expression for the
distance of closest approach of two hard ellipses with arbitrary orientation
in 2D. The strategy is to transform the ellipses into a circle and a new
ellipse by a scaling transformation. The relation between the position of a
point on the ellipse and the normal at that point allows the tangency
condition between the circle and ellipse to be written as a simple vector
equation with two unknowns, which may be solved analytically for the
distance between the centers. The solution requires the solution of a
quartic equation, whose single positive real root can be uniquely
determined. The final result for the distance is obtained by the inverse
scaling transformation. Explicit instructions for calculating the distance
are given in the Appendix. Our result may be useful in analytic and
numerical models of orientationally ordered systems.

\section{Appendix}

\subsection{Detailed calculation of the distance of closest approach}

We start by specifying all the quantities required for the calculation of $d$%
.

$a_{1}$ and $b_{1}$ are the lengths $(a_{1}>b_{1})$ of the major and minor
axes of ellipse $E_{1}$.

$a_{2}$ and $b_{2}$ are the lengths $(a_{2}>b_{2})$ of the major and minor
axes of ellipse $E_{2}$.

$e_{1}=\sqrt{1-\dfrac{b_{1}^{2}}{a_{1}^{2}}}$ is the eccentricity of ellipse
$E_{1}.$

$e_{2}=\sqrt{1-\dfrac{b_{2}^{2}}{a_{2}^{2}}}$ is the eccentricity of ellipse
$E_{2}.$

${\mathbf{\hat{k}}}_{1}\cdot\mathbf{\hat{d}}$ is the cosine of the angle
between the major axis $\mathbf{\hat{k}}_{1}$ of ellipse $E_{1}$ and the
direction $\mathbf{\hat{d}}$ of the line joining the centers.

${\mathbf{\hat{k}}}_{2}\cdot\mathbf{\hat{d}}$ is the cosine of the angle
between the major axis $\mathbf{\hat{k}}_{2}$ of ellipse $E_{2}$ and the
direction $\mathbf{\hat{d}}$ of the line joining the centers.

$\mathbf{\hat{k}}_{1}\cdot\mathbf{\hat{k}}_{2}$ is the cosine of the angle
between the major axis $\mathbf{\hat{k}}_{1}$ of ellipse $E_{1}$ and the
major axis $\mathbf{\hat{k}}_{2}$ of ellipse $E_{2}$.

The above quantities are specified in the statement of the problem.

The following quantities are derived from these.%
\begin{equation}
\eta =\dfrac{a_{1}}{b_{1}}-1.
\end{equation}%
In the coordinate system with the basis $(\mathbf{\hat{k}}_{1}+\mathbf{\hat{k%
}}_{2})/\sqrt{2+2\mathbf{\hat{k}}_{1}\cdot \mathbf{\hat{k}}_{2}}$ and $(%
\mathbf{\hat{k}}_{1}-\mathbf{\hat{k}}_{2})/\sqrt{2-2\mathbf{\hat{k}}%
_{1}\cdot \mathbf{\hat{k}}_{2}}$, the components of ${\mathbb{A}}^{\prime }$
are
\begin{equation}
{\mathbb{A}}_{11}^{\prime }=\frac{b_{1}^{2}}{b_{2}^{2}}(1+\frac{1}{2}(1+{%
\mathbf{\hat{k}}}_{1}\cdot {\mathbf{\hat{k}}}_{2})(\eta (2+\eta
)-e_{2}^{2}(1+\eta ({\mathbf{\hat{k}}}_{1}\cdot {\mathbf{\hat{k}}}%
_{2}))^{2})),
\end{equation}%
\begin{equation}
{\mathbb{A}}_{22}^{\prime }=\frac{b_{1}^{2}}{b_{2}^{2}}(1+\frac{1}{2}(1-{%
\mathbf{\hat{k}}}_{1}\cdot {\mathbf{\hat{k}}}_{2})(\eta (2+\eta
)-e_{2}^{2}(1-\eta ({\mathbf{\hat{k}}}_{1}\cdot {\mathbf{\hat{k}}}%
_{2}))^{2}))
\end{equation}%
and%
\begin{equation}
{\mathbb{A}}_{12}^{\prime }={\mathbb{A}}_{21}^{\prime }=\dfrac{b_{1}^{2}}{%
b_{2}^{2}}\frac{1}{2}\sqrt{1-({\mathbf{\hat{k}}}_{1}\cdot {\mathbf{\hat{k}}}%
_{2})^{2}}(\eta (2+\eta )+e_{2}^{2}(1-\eta ^{2}({\mathbf{\hat{k}}}_{1}\cdot {%
\mathbf{\hat{k}}}_{2})^{2})).
\end{equation}%
The eigenvalues of $\mathbb{A}^{\prime }$, in terms of these, are
\begin{equation}
\lambda _{+}=\dfrac{1}{2}({\mathbb{A}}_{11}^{\prime }+{\mathbb{A}}%
_{22}^{\prime })+\sqrt{\dfrac{1}{4}({\mathbb{A}}_{11}^{\prime }-{\mathbb{A}}%
_{22}^{\prime })^{2}+{\mathbb{A}}_{12}^{\prime 2}}
\end{equation}%
and
\begin{equation}
\lambda _{-}=\dfrac{1}{2}({\mathbb{A}}_{11}^{\prime }+{\mathbb{A}}%
_{22}^{\prime })-\sqrt{\dfrac{1}{4}({\mathbb{A}}_{11}^{\prime }-{\mathbb{A}}%
_{22}^{\prime })^{2}+{\mathbb{A}}_{12}^{\prime 2}}.
\end{equation}%
It follows that

\begin{equation}
b_{2}^{\prime }=\frac{1}{\sqrt{\lambda _{+}}}
\end{equation}%
and
\begin{equation}
a_{2}^{\prime }=\dfrac{1}{\sqrt{\lambda _{-}}}.
\end{equation}%
The eigenvectors are given by

\begin{equation}
\mathbf{\hat{k}}_{+}^{\prime }=\frac{1}{\sqrt{2}\sqrt{{{\mathbb{A}}%
_{12}^{\prime }}^{2}+(\lambda _{+}-{\mathbb{A}}_{11}^{\prime })^{2}}}\left(
\frac{{{\mathbb{A}}_{12}^{\prime }(\mathbf{\hat{k}}_{1}+\mathbf{\hat{k}}_{2})%
}}{\sqrt{1+\mathbf{\hat{k}}_{1}\cdot \mathbf{\hat{k}}_{2}}}+\frac{(\lambda
_{+}-{{\mathbb{A}}_{11}^{\prime })(\mathbf{\hat{k}}_{1}-\mathbf{\hat{k}}_{2})%
}}{\sqrt{1-\mathbf{\hat{k}}_{1}\cdot \mathbf{\hat{k}}_{2}}}\right)
\end{equation}%
and%
\begin{equation}
\mathbf{\hat{k}}_{-}^{\prime }=\frac{1}{\sqrt{2}\sqrt{{{\mathbb{A}}%
_{12}^{\prime }}^{2}+(\lambda _{+}-{\mathbb{A}}_{11}^{\prime })^{2}}}\left( -%
\frac{(\lambda _{+}-{{\mathbb{A}}_{11}^{\prime })(\mathbf{\hat{k}}_{1}+%
\mathbf{\hat{k}}_{2})}}{\sqrt{1+\mathbf{\hat{k}}_{1}\cdot \mathbf{\hat{k}}%
_{2}}}+\frac{{{\mathbb{A}}_{12}^{\prime }(\mathbf{\hat{k}}_{1}-\mathbf{\hat{k%
}}_{2})}}{\sqrt{1-\mathbf{\hat{k}}_{1}\cdot \mathbf{\hat{k}}_{2}}}\right) .
\end{equation}%
Then
\begin{align}
{\mathbf{\hat{k}}}_{+}^{\prime }\cdot \mathbf{\hat{d}}^{\prime }& =\cos \phi
=\frac{1}{\sqrt{2}\sqrt{{{\mathbb{A}}_{12}^{\prime }}^{2}+(\lambda _{+}-{%
\mathbb{A}}_{11}^{\prime })^{2}}\sqrt{1-e_{1}^{2}({\mathbf{\hat{k}}}%
_{1}\cdot \mathbf{\hat{d})}^{2}}}\times  \notag \\
& \left( \dfrac{{\mathbb{A}}_{12}^{\prime }}{\sqrt{1+\mathbf{\hat{k}}%
_{1}\cdot \mathbf{\hat{k}}_{2}}}(\dfrac{b_{1}}{a_{1}}({\mathbf{\hat{k}}}%
_{1}\cdot \mathbf{\hat{d})+}({\mathbf{\hat{k}}}_{2}\cdot \mathbf{\hat{d})+}%
\left( \dfrac{b_{1}}{a_{1}}-1\right) ({\mathbf{\hat{k}}}_{1}\cdot \mathbf{%
\hat{d})(}{\mathbf{\hat{k}}}_{1}\cdot \mathbf{\hat{k}}_{2}))+\right.  \notag
\\
& \left. \dfrac{(\lambda _{+}-{\mathbb{A}}_{11}^{\prime })}{\sqrt{1-\mathbf{%
\hat{k}}_{1}\cdot \mathbf{\hat{k}}_{2}}}(\dfrac{b_{1}}{a_{1}}({\mathbf{\hat{k%
}}}_{1}\cdot \mathbf{\hat{d})-(\hat{k}}_{2}\cdot \mathbf{\hat{d})-}\left(
\dfrac{b_{1}}{a_{1}}-1\right) ({\mathbf{\hat{k}}}_{1}\cdot \mathbf{\hat{d})(}%
{\mathbf{\hat{k}}}_{1}\cdot \mathbf{\hat{k}}_{2}))\right) .  \label{eq-cosf}
\end{align}

If $\mathbf{\hat{k}}_{1}=-\mathbf{\hat{k}}_{2}$, then $-\mathbf{\hat{k}}_{2}$
may be replaced by $+\mathbf{\hat{k}}_{2}$ without the loss of generality.
If $\mathbf{\hat{k}}_{1}=\mathbf{\hat{k}}_{2}$, care must be taken
evaluating the above expression. Letting $\mathbf{\hat{k}}_{1}\cdot \mathbf{%
\hat{k}}_{2}=\cos \theta $, in the limit as $\theta \rightarrow 0$ we find
that if ${\mathbb{A}}_{11}^{\prime }>{\mathbb{A}}_{22}^{\prime }$, then $%
(\lambda _{+}-{\mathbb{A}}_{11}^{\prime })\sim {\mathbb{A}}_{12}^{\prime 2}$%
, and
\begin{equation}
{\mathbf{\hat{k}}}_{+}^{\prime }\cdot \mathbf{\hat{d}}^{\prime }=\cos \phi =%
\frac{1}{\sqrt{1-e_{1}^{2}({\mathbf{\hat{k}}}_{1}\cdot \mathbf{\hat{d})}^{2}}%
}\dfrac{b_{1}}{a_{1}}({\mathbf{\hat{k}}}_{1}\cdot \mathbf{\hat{d})},
\label{eq-cosf2}
\end{equation}%
while if ${\mathbb{A}}_{11}^{\prime }<{\mathbb{A}}_{22}^{\prime }$, then
\begin{equation}
{\mathbf{\hat{k}}}_{+}^{\prime }\cdot \mathbf{\hat{d}}^{\prime }=\cos \phi =%
\frac{\sqrt{1-({\mathbf{\hat{k}}}_{1}\cdot \mathbf{\hat{d})}^{2}}}{\sqrt{%
1-e_{1}^{2}({\mathbf{\hat{k}}}_{1}\cdot \mathbf{\hat{d})}^{2}}}.
\label{eq-cosf3}
\end{equation}%
Next,
\begin{equation}
\delta =\frac{a_{2}^{\prime 2}}{b_{2}^{\prime 2}}-1.
\end{equation}%
If $\phi =\pi /2$ or if $\delta =0$, then $d^{\prime }=1+a_{2}^{\prime }$,
and Eq.\ (\ref{LAST}) can be evaluated directly. Otherwise,
\begin{equation}
\tan ^{2}\phi =\frac{1}{({\mathbf{\hat{k}}}_{+}^{\prime }\cdot \mathbf{\hat{d%
}}^{\prime })^{2}}-1,
\end{equation}%
\begin{subequations}
\begin{equation}
A=-\dfrac{1}{{b^{\prime }}_{2}^{2}}(1+\tan ^{2}\phi ),
\end{equation}%
\begin{equation}
B=-\dfrac{2}{{b_{2}^{\prime }}}(1+\tan ^{2}\phi +\delta ),
\end{equation}%
\begin{equation}
C=-\tan ^{2}\phi -(1+\delta )^{2}+\dfrac{1}{{b^{\prime }}_{2}^{2}}%
(1+(1+\delta )\tan ^{2}\phi ),
\end{equation}%
\begin{equation}
D=\dfrac{2}{{b_{2}^{\prime }}}(1+\tan ^{2}\phi )(1+\delta ),
\end{equation}%
\begin{equation}
E=(1+\tan ^{2}\phi +\delta )(1+\delta ),
\end{equation}%
\end{subequations}
\begin{subequations}
\begin{equation}
\alpha =-\dfrac{3B^{2}}{8A^{2}}+\dfrac{C}{A}
\end{equation}%
and
\begin{equation}
\beta =\dfrac{B^{3}}{8A^{3}}-\dfrac{BC}{2A^{2}}+\dfrac{D}{A}.
\end{equation}%
If $\beta \neq 0$, then
\begin{equation}
\gamma =\dfrac{-3B^{4}}{256A^{4}}+\dfrac{CB^{2}}{16A^{3}}-\dfrac{BD}{4A^{2}}+%
\dfrac{E}{A},
\end{equation}%
\begin{equation}
P=-\dfrac{\alpha ^{2}}{12}-\gamma ,
\end{equation}%
\begin{equation}
Q=-\dfrac{\alpha ^{3}}{108}+\dfrac{\alpha \gamma }{3}-\dfrac{\beta ^{2}}{8},
\end{equation}%
\begin{equation}
U=\left( -\dfrac{Q}{2}+\sqrt{\dfrac{Q^{2}}{4}+\dfrac{P^{3}}{27}}\right)
^{1/3}
\end{equation}%
and the principal values of the roots are taken throughout;
\begin{equation}
y=\left\{
\begin{array}{cc}
-\dfrac{5}{6}\alpha +U-\dfrac{P}{3U} & \text{ if }U\neq 0, \\
-\dfrac{5}{6}\alpha -Q^{1/3} & \text{if }U=0,%
\end{array}%
\right.
\end{equation}%
and
\begin{equation}
q=-\dfrac{B}{4A}+\dfrac{1}{2}\left( \sqrt{\alpha +2y}+\sqrt{-\left( 3\alpha
+2y+\dfrac{2\beta }{\sqrt{\alpha +2y}}\right) }\ \right) .
\end{equation}

If $\beta=0$, then
\end{subequations}
\begin{equation}
q=-\dfrac{B}{4A}+\sqrt{\frac{-\alpha+\sqrt{\alpha^{2}-4\gamma}}{2}},
\end{equation}

\begin{equation}
{d^{\prime }}=\sqrt{\dfrac{q^{2}-1}{\delta }\left( 1+\dfrac{b_{2}^{\prime
}(1+\delta )}{q}\right) ^{2}+\left( 1-\dfrac{q^{2}-1}{\delta }\right) \left(
1+\dfrac{b_{2}^{\prime }}{q}\right) ^{2}},
\end{equation}%
and finally
\begin{equation}
d=\frac{d^{\prime }}{\sqrt{1-e_{1}^{2}({\mathbf{\hat{k}}}_{1}\cdot \mathbf{%
\hat{d})}^{2}}}b_{1}.  \label{LAST}
\end{equation}

\subsection{Detailed calculation of the position of the point of contact}

To obtain the vector $\boldsymbol{r}_{c}$ from the center of Ellipse 1 to
the point of contact, we first need to compute ${\mathbf{\hat{n}}}^{\prime
}\,$ and then perform the inverse affine transformation. Explicitly, $%
\boldsymbol{r}_{c}={\mathbb{T}}^{-1}{\mathbf{\hat{n}}}^{\prime }$. The
components of ${\mathbf{\hat{n}}}^{\prime }$ can be defined though the inner
product with ${{\mathbf{\hat{k}}}}_{+}^{\prime }$ and ${\mathbf{\hat{k}}}%
_{-}^{\prime }$::%
\begin{eqnarray}
{{\mathbf{\hat{k}}}}_{+}^{\prime }\cdot {\mathbf{\hat{n}}}^{\prime } &=&\cos
\psi =sgn(\cos \phi )\sqrt{1-\frac{q^{2}-1}{\delta }}, \\
{\mathbf{\hat{k}}}_{-}^{\prime }\cdot {\mathbf{\hat{n}}}^{\prime } &=&\sin
\psi =sgn(\sin \phi )\sqrt{\frac{q^{2}-1}{\delta }}.
\end{eqnarray}%
where $sgn(x)$ gives the sign of $x$. The expression for $\cos \phi $ is
given in Eq. \ref{cosf}. \ $\sin \phi $ can be calculated similarly, and

\begin{align}
{\mathbf{\hat{k}}}_{-}^{\prime }\cdot \mathbf{\hat{d}}^{\prime }& =\sin \phi
=\frac{1}{\sqrt{2}\sqrt{{{\mathbb{A}}_{12}^{\prime }}^{2}+(\lambda _{+}-{%
\mathbb{A}}_{11}^{\prime })^{2}}\sqrt{1-e_{1}^{2}({\mathbf{\hat{k}}}%
_{1}\cdot \mathbf{\hat{d})}^{2}}}\times  \notag \\
& \left( -\dfrac{(\lambda _{+}-{\mathbb{A}}_{11}^{\prime })}{\sqrt{1+\mathbf{%
\hat{k}}_{1}\cdot \mathbf{\hat{k}}_{2}}}(\dfrac{b_{1}}{a_{1}}({\mathbf{\hat{k%
}}}_{1}\cdot \mathbf{\hat{d})+}({\mathbf{\hat{k}}}_{2}\cdot \mathbf{\hat{d})+%
}\left( \dfrac{b_{1}}{a_{1}}-1\right) ({\mathbf{\hat{k}}}_{1}\cdot \mathbf{%
\hat{d})(}{\mathbf{\hat{k}}}_{1}\cdot \mathbf{\hat{k}}_{2}))+\right.  \notag
\\
& \left. \dfrac{{\mathbb{A}}_{12}^{\prime }}{\sqrt{1-\mathbf{\hat{k}}%
_{1}\cdot \mathbf{\hat{k}}_{2}}}(\dfrac{b_{1}}{a_{1}}({\mathbf{\hat{k}}}%
_{1}\cdot \mathbf{\hat{d})-(\hat{k}}_{2}\cdot \mathbf{\hat{d})-}\left(
\dfrac{b_{1}}{a_{1}}-1\right) ({\mathbf{\hat{k}}}_{1}\cdot \mathbf{\hat{d})(}%
{\mathbf{\hat{k}}}_{1}\cdot \mathbf{\hat{k}}_{2}))\right) .
\end{align}

It follows that
\begin{equation}
{\mathbf{\hat{n}}}^{\prime }=\cos \psi {{\mathbf{\hat{k}}}}_{+}^{\prime
}+\sin \psi {\mathbf{\hat{k}}}_{-}^{\prime }.
\end{equation}%
Writing ${{\mathbf{\hat{k}}}}_{+}^{\prime }$ and ${\mathbf{\hat{k}}}%
_{-}^{\prime }\,$in terms of ${\mathbf{\hat{k}}}_{1}$ and ${\mathbf{\hat{k}}}%
_{2}$
\begin{equation}
{{\mathbf{\hat{k}}}}_{+}^{\prime }=\cos \gamma \frac{(\mathbf{\hat{k}}_{1}+%
\mathbf{\hat{k}}_{2})}{\sqrt{2}\sqrt{1+\mathbf{\hat{k}}_{1}\cdot \mathbf{%
\hat{k}}_{2}}}+\sin \gamma \frac{(\mathbf{\hat{k}}_{1}-\mathbf{\hat{k}}_{2})%
}{\sqrt{2}\sqrt{1-\mathbf{\hat{k}}_{1}\cdot \mathbf{\hat{k}}_{2}}},
\end{equation}%
\begin{equation}
{{\mathbf{\hat{k}}}}_{-}^{\prime }=-\sin \gamma \frac{(\mathbf{\hat{k}}_{1}+%
\mathbf{\hat{k}}_{2})}{\sqrt{2}\sqrt{1+\mathbf{\hat{k}}_{1}\cdot \mathbf{%
\hat{k}}_{2}}}+\cos \gamma \frac{(\mathbf{\hat{k}}_{1}-\mathbf{\hat{k}}_{2})%
}{\sqrt{2}\sqrt{1-\mathbf{\hat{k}}_{1}\cdot \mathbf{\hat{k}}_{2}}},
\end{equation}%
where%
\begin{equation}
\cos \gamma ={{\mathbf{\hat{k}}}}_{+}^{\prime }\cdot \frac{(\mathbf{\hat{k}}%
_{1}+\mathbf{\hat{k}}_{2})}{\sqrt{2}\sqrt{1+\mathbf{\hat{k}}_{1}\cdot
\mathbf{\hat{k}}_{2}}},
\end{equation}%
and%
\begin{equation}
\sin \gamma ={{\mathbf{\hat{k}}}}_{+}^{\prime }\cdot \frac{(\mathbf{\hat{k}}%
_{1}-\mathbf{\hat{k}}_{2})}{\sqrt{2}\sqrt{1-\mathbf{\hat{k}}_{1}\cdot
\mathbf{\hat{k}}_{2}}}.
\end{equation}

The angle $\gamma $ can be calculated explicitly, since ${{\mathbf{\hat{k}}}}%
_{+}^{\prime }$ and ${\mathbf{\hat{k}}}_{-}^{\prime }$ \ are known. \
Substitution gives ${\mathbf{\hat{n}}}^{\prime }$ in terms of ${\mathbf{\hat{%
k}}}_{1}$ and ${\mathbf{\hat{k}}}_{2}$\ and gives

\begin{equation}
{\mathbf{\hat{n}}}^{\prime }=\cos (\psi +\gamma )\frac{(\mathbf{\hat{k}}_{1}+%
\mathbf{\hat{k}}_{2})}{\sqrt{2}\sqrt{1+\mathbf{\hat{k}}_{1}\cdot \mathbf{%
\hat{k}}_{2}}}+\sin (\psi +\gamma )\frac{(\mathbf{\hat{k}}_{1}-\mathbf{\hat{k%
}}_{2})}{\sqrt{2}\sqrt{1-\mathbf{\hat{k}}_{1}\cdot \mathbf{\hat{k}}_{2}}}.
\end{equation}

Finally, after the transformaton ${\mathbb{T}}^{-1}$, the contact point $%
\boldsymbol{r}_{c}$ is given by a linear combination of $\mathbf{\hat{k}}%
_{1} $ and $\mathbf{\hat{k}}_{2}:$

\begin{eqnarray}
\boldsymbol{r}_{c} &=&{\mathbb{T}}^{-1}{\mathbf{\hat{n}}}^{\prime }={b{_{1}}%
\left( {\mathbf{I}}+\eta {\mathbf{\hat{k}}}_{1}{\mathbf{\hat{k}}}_{1}\right)
\mathbf{\hat{n}}}^{\prime }  \notag \\
&=&[(a_{1}+(a_{1}-b_{1})\mathbf{\hat{k}}_{1}\cdot \mathbf{\hat{k}}_{2})\frac{%
\cos (\psi +\gamma )}{\sqrt{2}\sqrt{1+\mathbf{\hat{k}}_{1}\cdot \mathbf{\hat{%
k}}_{2}}}+(a_{1}-(a_{1}-b_{1})\mathbf{\hat{k}}_{1}\cdot \mathbf{\hat{k}}_{2})%
\frac{\sin (\psi +\gamma )}{\sqrt{2}\sqrt{1-\mathbf{\hat{k}}_{1}\cdot
\mathbf{\hat{k}}_{2}}}]\mathbf{\hat{k}}_{1}+  \notag \\
&&\lbrack b_{1}\cos (\psi +\gamma )\frac{1}{\sqrt{2}\sqrt{1+\mathbf{\hat{k}}%
_{1}\cdot \mathbf{\hat{k}}_{2}}}-b_{1}\sin (\psi +\gamma )\frac{\mathbf{1}}{%
\sqrt{2}\sqrt{1-\mathbf{\hat{k}}_{1}\cdot \mathbf{\hat{k}}_{2}}}]\mathbf{%
\hat{k}}_{2}.
\end{eqnarray}

There are some special cases, when either $q$ is not given or $\phi $ is not
well defined \bigskip We treat these separately below.

\begin{enumerate}
\item If $\delta =0$ or $\phi =\pi /2$, then $\cos \psi $ and $\sin \psi $
are not required, since that ${\mathbf{\hat{n}}}^{\prime }=$ $\mathbf{\hat{d}%
}^{\prime }$and
\begin{equation}
\boldsymbol{r}_{c}={\mathbb{T}}^{-1}\mathbf{\hat{d}}^{\prime }=\frac{{%
\mathbb{T}}^{-1}{\mathbb{T}}\mathbf{\hat{d}}}{\left\vert {\mathbb{T}}\mathbf{%
\hat{d}}\right\vert }=\frac{\mathbf{\hat{d}}}{\frac{1}{b_{1}}\sqrt{%
1-e_{1}^{2}({\mathbf{\hat{k}}}_{1}\cdot \mathbf{\hat{d})}^{2}}}.
\end{equation}

\item If ${{\mathbf{\hat{k}}}}_{1}={{\mathbf{\hat{k}}}}_{2}$, care must to
be taken in evaluating the angle $\phi $. \ Since ${{\mathbf{\hat{k}}}}_{1}={%
{\mathbf{\hat{k}}}}_{2}$, these vectors do not span the space, and a new
vector ${\mathbf{\hat{k}}}_{1}^{\perp }$, perpendicular to ${{\mathbf{\hat{k}%
}}}_{1}$, needs to be introduced. $\,.$

\begin{enumerate}
\item If $A_{11}>A_{22},$ then ${{\mathbf{\hat{k}}}}_{+}^{\prime }$ $={%
\mathbf{\hat{k}}}_{1}$, and ${\mathbf{\hat{k}}}_{-}^{\prime }={\mathbf{\hat{k%
}}}_{1}^{\perp }$, and $\cos \phi $ and $\sin \phi $ are given by%
\begin{eqnarray}
{\mathbf{\hat{k}}}_{+}^{\prime }\cdot \mathbf{\hat{d}}^{\prime } &=&\cos
\phi =\frac{1}{\sqrt{1-e_{1}^{2}({\mathbf{\hat{k}}}_{1}\cdot \mathbf{\hat{d})%
}^{2}}}\dfrac{b_{1}}{a_{1}}({\mathbf{\hat{k}}}_{1}\cdot \mathbf{\hat{d}),} \\
{\mathbf{\hat{k}}}_{-}^{\prime }\cdot \mathbf{\hat{d}}^{\prime } &=&\sin
\phi =\frac{1}{\sqrt{1-e_{1}^{2}({\mathbf{\hat{k}}}_{1}\cdot \mathbf{\hat{d})%
}^{2}}}({\mathbf{\hat{k}}}_{1}^{\perp }\cdot \mathbf{\hat{d})},
\end{eqnarray}

Here ${\mathbf{\hat{n}}}^{\prime }=\cos \psi {{\mathbf{\hat{k}}}}_{1}+\sin
\psi {\mathbf{\hat{k}}}_{1}^{\perp }$, and the $\boldsymbol{r}_{c}$ is given
by%
\begin{eqnarray}
\boldsymbol{r}_{c} &=&{\mathbb{T}}^{-1}\mathbf{\hat{d}}^{\prime }={b_{1}}%
\left( {\mathbf{I}}+\eta {\mathbf{\hat{k}}}_{1}{\mathbf{\hat{k}}}_{1}\right)
(\cos \psi {{\mathbf{\hat{k}}}}_{1}+\sin \psi {\mathbf{\hat{k}}}_{1}^{\perp
})  \notag \\
&=&a_{1}\cos \psi {\mathbf{\hat{k}}}_{1}+b_{1}\sin \psi {\mathbf{\hat{k}}}%
_{1}^{\perp }.
\end{eqnarray}

\item If $A_{11}<A_{22},$ ${{\mathbf{\hat{k}}}}_{+}^{\prime }$ $={\mathbf{%
\hat{k}}}_{1}^{\perp }$, and ${\mathbf{\hat{k}}}_{-}^{\prime }={{\mathbf{%
\hat{k}}}}_{1}$, then
\begin{eqnarray}
{\mathbf{\hat{k}}}_{+}^{\prime }\cdot \mathbf{\hat{d}}^{\prime } &=&\cos
\phi =\frac{1}{\sqrt{1-e_{1}^{2}({\mathbf{\hat{k}}}_{1}\cdot \mathbf{\hat{d})%
}^{2}}}({\mathbf{\hat{k}}}_{1}^{\perp }\cdot \mathbf{\hat{d})}, \\
{\mathbf{\hat{k}}}_{-}^{\prime }\cdot \mathbf{\hat{d}}^{\prime } &=&\sin
\phi =\frac{1}{\sqrt{1-e_{1}^{2}({\mathbf{\hat{k}}}_{1}\cdot \mathbf{\hat{d})%
}^{2}}}\dfrac{b_{1}}{a_{1}}({\mathbf{\hat{k}}}_{1}\cdot \mathbf{\hat{d})},
\end{eqnarray}

and ${\mathbf{\hat{n}}}^{\prime }=\cos \psi {\mathbf{\hat{k}}}_{1}^{\perp
}+\sin \psi {\mathbf{\hat{k}}}_{1}$ \ We then have
\begin{eqnarray}
\boldsymbol{r}_{c} &=&{\mathbb{T}}^{-1}\mathbf{\hat{d}}^{\prime }={b_{1}}%
\left( {\mathbf{I}}+\eta {\mathbf{\hat{k}}}_{1}{\mathbf{\hat{k}}}_{1}\right)
(\cos \psi {\mathbf{\hat{k}}}_{1}^{\perp }+\sin \psi {\mathbf{\hat{k}}}_{1})
\notag \\
&=&b_{1}\cos \psi {\mathbf{\hat{k}}}_{1}^{\perp }+a_{1}\sin \psi {\mathbf{%
\hat{k}}}_{1}.
\end{eqnarray}
\end{enumerate}
\end{enumerate}

\section{Acknowledgments}

One of us (P.P-M.) acknowledges useful discussions with D. Frenkel and B.
Mulder. This work was supported in part by the NSF under DMS 0440299.

\end{document}